\documentclass[aps]{revtex4}
\usepackage{float}
\usepackage{graphicx}
\usepackage{caption}
\usepackage{amsmath}
\usepackage{array}
\usepackage{bm}
\usepackage{amssymb}
\usepackage{amsfonts}
\usepackage{amssymb}
\usepackage{color}

\begin{document}
\title{Damping and polarization rates in near equilibrium state}
\author{Ziyue Wang}
\author{Pengfei Zhuang}
\affiliation{Physics Department, Tsinghua University, Beijing 100084, China}
\date{\today}
\begin{abstract}
The collision terms in spin transport theory are analyzed in Kadanoff-Baym formalism for systems close to equilibrium. The non-equilibrium fluctuations in spin distribution include both damping and polarization, with the latter arising from the exchange between orbital and spin angular momenta. The damping and polarization rates or the relaxation times are expressed in terms of various Dirac components of the self-energy. Unlike the usually used Anderson-Witting relaxation time approximation assuming a single time scale for different degrees of freedom, the polarization effect is induced by the thermal vorticity and its time scale of thermalization is different from the damping. The numerical calculation in the Nambu--Jona-Lasinio model shows that, charge is thermalized earlier and spin is thermalized later.  
\end{abstract}
\maketitle

\section{Introduction}
\label {s1}
The Quark-Gluon Plasma (QGP) created in relativistic heavy-ion collisions is a special laboratory to study strong interaction at high temperature and density~\cite{Velkovska:2004ff, Adams:2005dq}. In non-central heavy ion collisions, a large global angular momentum is produced, and the spin of hadrons emitted is aligned with its direction~\cite{STAR:2017ckg, Adam:2018ivw}. The magnitude of the global polarization of $\Lambda$ hyperons can be well described by models based on relativistic hydrodynamics, assuming local equilibrium of the spin degrees of freedom~\cite{Becattini:2013vja, Becattini:2016gvu, Karpenko:2016jyx, Pang:2016igs, Xie:2017upb}. However, the data on the sophisticated polarization structure, in particular the longitudinal polarization at different azimuthal angles~\cite{Adam:2019srw}, disagree with current model calculations~\cite{Becattini:2015ska, Becattini:2017gcx, Xia:2018tes}. Theoretical attempts to solve this discrepancy have been devoted to including off-equilibrium effect in the evolution of spin. In a macroscopic sense, spin-hydrodynamical models are constructed based on conservation laws and the second law of thermodynamics. In the microscopic sense, spin transport theory~\cite{Son:2012bg, Son:2012wh, Son:2012zy, Stephanov:2012ki, Pu:2010as, Chen:2012ca, Hidaka:2016yjf, Huang:2018wdl, Liu:2018xip, Lin:2019ytz, Hattori:2019ahi, Wang:2019moi, Gao:2019znl, Weickgenannt:2019dks, Liu:2020flb} based on the Wigner-function formalism\cite{DeGroot:1980dk} has been developed to describe spin related anomalous transport phenomena in heavy ion collisions. 

The collision effect is crucial for off-equilibrium systems, and plays a dominant role in the evolution towards thermal equilibrium. Based on the Schwinger-Keldysh formalism~\cite{Martin:1959jp, Keldysh:1964ud}, the spin transport theory has been extended from the free streaming scenario to including collisions~\cite{Yang:2020hri}. For the evolution of spin, there exist both damping and polarization in collisions~\cite{Zhang:2019xya, Weickgenannt:2020aaf}. The polarization can be captured through semi-classical expansion at order $\mathcal{O}(\hbar)$. Based on the framework developed in Ref.\cite{Yang:2020hri}, the spin-diffusion term for massive quarks is calculated up to the leading logarithmic order~\cite{Yang:2020hri}, a new polarization direction is discovered in an off-equilibrium and weakly coupled electrodynamics plasma~\cite{Hou:2020mqp}, and the global equilibrium spin distribution function is obtained through detailed balance principle~\cite{Wang:2020pej}.

The full collision terms in spin transport can be derived order by order in $\hbar$, yet the result is complicated. The relaxation time approximation (RTA) is a widely employed simplification to the collision kernel in transport equations. Due to its simplicity, it is useful in deriving the kinetic coefficients~\cite{Jaiswal:2014isa, Czajka:2017wdo} and dissipative hydrodynamic equations~\cite{Jaiswal:2013vta} as well as their exact solutions~\cite{Denicol:2014xca,Florkowski:2014sfa}. RTA in spin transport has also been used as a starting point to derive the spin-hydrodynamical equations~\cite{Bhadury:2020cop}. However, the normally used RTA is based on the assumption that all microscopic interactions share the same timescale, ignoring the interaction mechanism of the microscopic constituents.    

In this work we study the RTA of spin in Wigner function formalism. We derive the relaxation equations directly from the Kadanoff-Baym equations for a system close to equilibrium. We will point out that besides the damping effect the RTA contains a polarization effect. The damping rate and polarization rate can be calculated explicitly after specifying the interaction. We adopt the NJL model as an example to calculate the ratio between the two rates and compare the result with the usually used RTA~\cite{Hakim:1992}.

The paper is arranged as follows, in Sec.\ref{s2} we review the key steps in deriving the transport equations in the Kadanoff-Baym formalism. In Sec.\ref{s3} the Anderson-Witting RTA (A-W RTA) for fermion Wigner function is discussed, and the polarization effect in the transport equation for the axial-vector component is emphasized. In Sec.\ref{s4} the transport equations derived in the Kadanoff-Baym formalism are simplified for systems close to equilibrium, and the damping and polarization rates are expressed in terms of the fermion self-energies. In Sec.\ref{s5} the ratio between the polarization and damping rates is calculate as a concrete example in the NJL model. Eventually, we make concluding remarks and outlook in Sec.\ref{s6}. The details on the spin decomposition of the collision terms is presented in the Appendix. 

\section{Transport and constraint equations}
\label{s2}
In this section, we review the basic steps of deriving the spin transport equation with collision terms. Starting from the Wigner transformation applied to contour Green's function,
\begin{equation}
S_{\alpha\beta}^{<,>}(x,p)=\int d^4 y e^{ip\cdot y/\hbar}\tilde{S}_{\alpha\beta}^{<,>}(x+y/2,x-y/2),
\end{equation}
where $\tilde{S}_{\alpha\beta}^{<}(x+y/2,x-y/2)=\langle\bar{\psi}_\beta(x-y/2)\psi_\alpha(x+y/2)\rangle$ and $\tilde{S}_{\alpha\beta}^{>}(x+y/2,x-y/2)=\langle\psi_\alpha(x+y/2)\bar{\psi}_\beta(x-y/2)\rangle$ are lessor and greater fermion propagators, respectively. The Wigner transformation of the Dyson-Schwinger equations for the lessor and greater propagators gives the Kadanoff-Baym equations~\cite{Yang:2020hri}, and the sum and difference of these equations lead to the constraint and transport equations for the fermion Wigner function $W(x,p)\equiv S^{<}(x,p)$, 
\begin{eqnarray}
\label{KB}
\left\{(\gamma^\mu p_\mu-m),S^<\right\}+\frac{i\hbar}{2}\left[\gamma^\mu,\partial_\mu S^<\right] &=&\frac{i\hbar}{2}C,\nonumber\\
\left[(\gamma^\mu p_\mu-m),S^<\right]+\frac{i\hbar}{2}\left\{\gamma^\mu,\partial_\mu S^<\right\} &=&\frac{i\hbar}{2}D
\end{eqnarray}
with collision terms $C(x,p)$ and $D(x,p)$,
\begin{eqnarray}
\label{CD}
C &=& \left[\Sigma^<,S^>\right]_\star-\left[\Sigma^>,S^<\right]_\star,\nonumber\\
D &=& \left\{\Sigma^<,S^>\right\}_\star-\left\{\Sigma^>,S^<\right\}_\star,
\end{eqnarray}
where $m$ is the fermion mass, and $\Sigma^{<,>}(x,p)$ are the lessor and greater self-energies. Since a scattering process involves only $\Sigma^{<,>}$, we have dropped here the real parts of the retarded and advanced self-energies and propagators. The commutators for any two functions $A(x,p)$ and $B(x,p)$ are defined as $\{A,B\}\equiv AB+BA$, $[A,B]\equiv AB-BA$, $\{A,B\}_\star\equiv A\star B+B\star A$ and $[A,B]_\star\equiv A\star B-B\star A$. The star product is generated from the Wigner transformation and stands for the shorthand notation $A\star B=AB+i\hbar/2[AB]_{\text{PB}}+\mathcal{O}(\hbar^2)$. The Poisson bracket $[AB]_{\text{PB}}\equiv(\partial_p^\mu A)(\partial_\mu B)-(\partial_\mu A)(\partial_p^\mu B)$ contains coordinate and momentum derivatives $\partial_\mu$ and $\partial^p_\mu$. 

The Wigner function is a matrix in Dirac space and not a hermitian matrix, $W^+=\gamma_0 W \gamma_0\neq W$. To derive kinetic equations for physical distributions, one should perform spin decomposition for the Wigner function, self-energies and collision terms with the Clifford algebra, 
\begin{eqnarray}
\label{spin}
S^< &=& F+iP\gamma_5+V^\mu\gamma_\mu+A^\mu\gamma_5\gamma_\mu+\frac{1}{2}S^{\mu\nu}\sigma_{\mu\nu},\nonumber\\
S^> &=& \bar{F}+i\bar{P}\gamma_5+\bar{V}^\mu\gamma_\mu+\bar{A}^\mu\gamma_5\gamma_\mu+\frac{1}{2}\bar{S}^{\mu\nu}\sigma_{\mu\nu},\nonumber\\
\Sigma^< &=& \Sigma_F+i\Sigma_P\gamma_5+\Sigma_V^\mu\gamma_\mu+\Sigma_A^\mu\gamma_5\gamma_\mu+\frac{1}{2}\Sigma_S^{\mu\nu}\sigma_{\mu\nu},\nonumber\\
\Sigma^> &=& \bar{\Sigma}_F+i\bar{\Sigma}_P\gamma_5+\bar{\Sigma}_V^\mu\gamma_\mu+\bar{\Sigma}_A^\mu\gamma_5\gamma_\mu+\frac{1}{2}\bar{\Sigma}_S^{\mu\nu}\sigma_{\mu\nu},\nonumber\\
C  &=& C_F+iC_P\gamma_5+C_V^\mu\gamma_\mu+C_A^\mu\gamma_5\gamma_\mu+\frac{1}{2}C_S^{\mu\nu}\sigma_{\mu\nu},\nonumber\\
D  &=& D_F+iD_P\gamma_5+D_V^\mu\gamma_\mu+D_A^\mu\gamma_5\gamma_\mu+\frac{1}{2}D_S^{\mu\nu}\sigma_{\mu\nu}.
\end{eqnarray}

From the sum and difference of the Kadanoff-Baym equations (\ref{KB}) as well as the spin decomposition (\ref{spin}), one derives the kinetic equations for the spin components of the Wigner function, 
\begin{eqnarray}
\label{constraint}
&&p_\mu V^\mu-m S=\frac{i\hbar}{4}C_F,\nonumber\\
&&2m P+ \hbar\partial_\mu A^\mu=-\frac{i\hbar}{2}C_P,\nonumber\\
&&2p^\mu F-2m V^\mu-\hbar\partial_\nu S^{\nu\mu}=\frac{i\hbar}{2}C_V^\mu,\nonumber\\
&&\hbar\partial^\mu P-\epsilon^{\mu\nu\rho\sigma}p_\sigma S_{\nu\rho}-2m A_\mu=\frac{i\hbar}{2}C_A^\mu,\nonumber\\
&&\hbar\partial^{[\mu} V^{\nu]}-2\epsilon^{\rho\sigma\mu\nu}p_\rho A_\sigma-2m S^{\mu\nu}=\frac{i\hbar}{2}C_S^{\mu\nu}
\end{eqnarray}
and 
\begin{eqnarray}
\label{transport}
&&\partial_\mu V^\mu=\frac{1}{2}D_F,\nonumber\\
&&2p_\mu A^\mu=\frac{\hbar}{2}D_P,\nonumber\\
&&2p_\nu S^{\nu\mu}+ \hbar\partial^\mu F=\frac{\hbar}{2}D_V^\mu,\nonumber\\
&&2p^\mu P+\frac{\hbar}{2}\epsilon^{\mu\nu\rho\sigma}\partial_\sigma S_{\nu\rho}=-\frac{\hbar}{2}D_A^\mu,\nonumber\\
&&2p^{[\mu} V^{\nu]}+\hbar\epsilon^{\mu\nu\rho\sigma}\partial_\rho A_\sigma=-\frac{\hbar}{2}D_S^{\mu\nu}.
\end{eqnarray}
The physics of the spin components can be extracted from the conservation laws in phase space~\cite{Vasak:1987um}. Considering the energy-momentum conservation and total angular momentum conservation, the vector and axial-vector components $V_\mu(x,p)$ and $A_\mu(x,p)$ are fermion number and spin currents. 

To perturbatively solve the kinetic equations (\ref{constraint}) and (\ref{transport}), one usually takes semi-classical expansion, 
\begin{eqnarray}
W &=& W^{(0)}+\hbar W^{(1)}+\hbar^2 W^{(2)}+\cdots,\nonumber\\
\Sigma &=& \Sigma^{(0)}+\hbar\Sigma^{(1)}+\hbar^2\Sigma^{(2)}+\cdot,\nonumber\\
C &=& C^{(0)}+\hbar C^{(1)}+\hbar^2 C^{(2)}+\cdots,\nonumber\\
D &=& D^{(0)}+\hbar D^{(1)}+\hbar^2 D^{(2)}+\cdots
\end{eqnarray}
for the Wigner function, self-energies and collision terms. At classical level and first order $\mathcal O(\hbar)$, the 16 spin components are not all independent~\cite{Vasak:1987um}. We can choose the independent ones as $V_\mu$ and $A_\mu$, and the other 8 scalar, pseudo-scalar and tensor components $F, P$ and $S_{\mu\nu}$ can be expressed in terms of them, 
\begin{eqnarray}
\label{relation}
F^{(0)}&=&\frac{p^\mu}{m}V^{(0)}_\mu,\nonumber\\
F^{(1)}&=&\frac{p^\mu}{m}V^{(1)}_\mu-\frac{i}{4m}C_F^{(0)},\nonumber \\
P^{(0)}&=&0,\nonumber\\
P^{(1)}&=&-\frac{1}{2m}\partial^\mu A^{(0)}_\mu-\frac{i}{4m}C_P^{(0)},\nonumber  \\
S_{\mu\nu}^{(0)}&=&-\frac{1}{m}\epsilon_{\rho\sigma\mu\nu}p^\rho A^{(0)\sigma}, \nonumber\\
S^{(1)}_{\mu\nu}&=&\frac{1}{2m}\partial_{[\mu} V^{(0)}_{\nu]}-\frac{1}{m}\epsilon_{\rho\sigma\mu\nu}p^\rho A^{(1)\sigma}-\frac{i}{4m}C^{(0)}_{S\mu\nu}.
\end{eqnarray}
The constraint equations $p^\mu A^{(0)}_\mu=0$ and $p_{[\mu}V^{(0)}_{\nu]}\equiv p_\mu V_\nu^{(0)}-p_\nu V_\mu^{(0)}=0$ at classical level and the similar constraints at first order indicate that, only three Lorentz components of $A_\mu$ and one Lorentz component of $V_\mu$ are independent. We can take them as ${\bf A}$ and $V_0$, corresponding to the number density and spin density. In order to keep the description covariant and symmetric, we in the following derive the transport equations for $V_\mu$ and $A_\mu$, but keep in mind that $V_\mu$ and $A_\mu$ contain only four independent degrees of freedom in total. 

The classical components are all on the mass shell and their transport equations are in the Boltzmann type, 
\begin{eqnarray}
&& (p^2-m^2)V^{(0)}_\mu=0,\nonumber\\
&& (p^2-m^2)A^{(0)}_\mu=0,\nonumber\\
&& p^\nu\partial_\nu V^{(0)}_\mu=\frac{m}{2}D_{V\mu}^{(0)}+\frac{ip^\nu}{2} C_{S\nu\mu}^{(0)},\nonumber\\
&& p^\nu\partial_\nu A^{(0)}_\mu=\frac{m}{2}D_{A\mu}^{(0)}-\frac{ip_\mu}{2} C_P^{(0)}.
\end{eqnarray}
With the collision terms expressed in terms of the self-energies shown in Appendix \ref{a1}, the classical transport equations are explicitly written as  
\begin{eqnarray}
\label{TVA0}
p^\nu\partial_\nu V^{(0)}_\mu &=& m\overline{\Sigma_F V_{\mu}}^{(0)} +p_\nu\overline{\Sigma_V^\nu V_{\mu}}^{(0)}
+\frac{m}{2}\epsilon_{\alpha\beta\lambda\mu}\overline{\Sigma_S^{\alpha\beta}A^\lambda}^{(0)}
-\frac{p_\nu}{m}\epsilon_{\alpha\mu\beta\lambda}p^\beta\overline{\Sigma_S^{\alpha\nu}A^{\lambda}}^{(0)}-p_\mu\overline{\Sigma_{A}^{\nu}A_{\nu}}^{(0)},\nonumber\\
p^\nu\partial_\nu A^{(0)}_\mu &=& m\overline{\Sigma_FA_\mu}^{(0)} +p_\nu\overline{\Sigma_V^\nu A_\mu}^{(0)} +\frac{m}{2}\epsilon_{\alpha\beta\lambda\mu}\overline{\Sigma_S^{\alpha\beta}V^{\lambda}}^{(0)}
+p^\nu\overline{\Sigma_{A\mu} V_\nu}^{(0)}-p_\mu\overline{\Sigma_A^{\nu}V_\nu}^{(0)},
\end{eqnarray}
where the hat operators are defined as $\overline{AB}^{(0)}=\overline{A^{(0)}B^{(0)}}$ and $\overline{AB}=\overline AB-A\overline B$. 

Since the spin polarization is in general a quantum effect, it is crucial to investigate transport equations at the first order of $\hbar$, especially for the transport equation for $A_\mu^{(1)}$. Considering the classical relations among the spin components shown in Eq.(\ref{relation}), we obtain the modified on-shell conditions and transport equations, 
\begin{eqnarray}
(p^2-m^2)V^{(1)}_\mu &=& \frac{ip_\mu}{4}C_F^{(0)}+\frac{im}{4}C_{V\mu}^{(0)},\nonumber\\
(p^2-m^2)A_\mu^{(1)} &=& \frac{p_\mu}{4}D_P^{(0)}-\frac{i}{8}\epsilon_{\mu\alpha\beta\gamma}p^\alpha C_{S}^{(0)\beta\gamma}+\frac{i}{4}mC_{A\mu}^{(0)},\nonumber\\
(p\cdot\partial)V_\mu^{(1)} &=& \frac{m}{2}D_{V\mu}^{(1)}+\frac{ip^\nu}{2} C_{S\nu\mu}^{(1)}+\frac{i}{4}\partial_\mu C_F^{(0)},\nonumber\\
(p\cdot\partial)A^{(1)}_\mu &=& \frac{m}{2}D_{A\mu}^{(1)}-\frac{ip_\mu}{2} C_P^{(1)}-\frac{i}{8}\epsilon_{\mu\sigma\nu\rho}\partial^\sigma C_{S}^{(0)\nu\rho}.
\end{eqnarray}
Taking the expressions of the collision terms in terms of the self-energies shown in Appendix \ref{a1}, the modified on-shell conditions become
\begin{eqnarray}
(p^2-m^2)V^{(1)}_\mu &=&-\frac{m}{2}\overline{\Sigma_P A_\mu}^{(0)} -\frac{1}{2}\epsilon_{\mu\nu\alpha\beta}p^\alpha\overline{\Sigma_{V}^\nu A^\beta}^{(0)}
-\frac{m}{2}\overline{\Sigma_{S\mu\nu}V^\nu}^{(0)},\nonumber\\
(p^2-m^2)A_\mu^{(1)} &=&\frac{p_\mu p^\nu}{2m}\overline{\Sigma_PV_\nu}^{(0)}-\frac{m}{2}\overline{\Sigma_P V_\mu}^{(0)}+\epsilon_{\mu\alpha\beta\gamma}p^\alpha\overline{\Sigma_A^\beta A^\gamma}^{(0)},
\end{eqnarray}
which indicate that in quantum case the spin components $V_\mu$ and $A_\mu$ and in turn the others are no longer on the mass-shell. 

When quantum fluctuations are included, the transport equations for $V^{(1)}_\mu$ and $A^{(1)}_\mu$ with collision terms in terms of the self-energies become much more complicated in comparison with the classical ones,  
\begin{eqnarray}
\label{TVA1}
(p\cdot\partial)V^{(1)}_\mu 
&=& m\overline{\Sigma_FV_\mu}^{(1)} +p_\nu\overline{\Sigma_V^\nu V_\mu}^{(1)} +\frac{m}{2}\epsilon_{\alpha\beta\lambda\mu}\overline{\Sigma_{S}^{\alpha\beta}A^\lambda}^{(1)}
-\frac{p_\nu}{m}\epsilon_{\alpha\mu\beta\lambda}p^\beta\overline{\Sigma_S^{\alpha\nu}A^\lambda}^{(1)}-p_\mu\overline{\Sigma_A^\nu A_\nu}^{(1)}\nonumber\\
&&
+\frac{p^\nu}{2m}\left[\overline{\Sigma_{S\mu\nu}\left(p^\alpha V_\alpha\right)}^{(0)}\right]_{\text{PB}}
-\frac{m}{2}\left[\overline{\Sigma_{S\mu\nu}V^\nu}^{(0)}\right]_{\text{PB}}
+\frac{1}{2}\epsilon_{\mu\nu\alpha\beta}\left[\overline{\Sigma_A^\alpha\left(p^\nu V^\beta\right)}^{(0)}\right]_{\text{PB}}\nonumber\\
&&
-\frac{p_\nu}{2m}\overline{\Sigma_{S\alpha\mu}\partial^{[\alpha}V^\nu]}^{(0)}
+\frac{p_\nu}{2m}\overline{\Sigma_S^{\alpha\nu}\partial_{[\alpha}V_{\mu]}}^{(0)}
+\frac{1}{2}\epsilon_{\beta\nu\lambda\mu}\overline{\Sigma_A^\beta\partial^\nu V^\lambda}^{(0)}
+\frac{1}{2}\epsilon_{\mu\nu\alpha\beta}(\partial^\alpha\overline{\Sigma_V^\nu)A^\beta}^{(0)}\nonumber\\
&&
-\frac{p_\mu}{2m}\overline{(\partial^\nu\Sigma_P)A_\nu}^{(0)}
-\frac{p^\nu}{2m}\overline{(\partial_\nu\Sigma_P)A_\mu}^{(0)}
+\frac{p^\nu}{2m}\epsilon_{\mu\nu\alpha\beta}\overline{(\partial^\alpha\Sigma_F)A^\beta}^{(0)},\nonumber\\
(p\cdot\partial)A_\mu^{(1)}
&=&
m\overline{\Sigma_FA_{\mu}}^{(1)}+p_\nu\overline{\Sigma_V^\nu A_\mu}^{(1)}+\frac{m}{2}\epsilon_{\alpha\beta\lambda\mu}\overline{\Sigma_S^{\alpha\beta}V^\lambda}^{(1)}
+p^\nu\overline{\Sigma_\mu V_\nu}^{(1)}-p_\mu\overline{\Sigma_A^\nu V_\nu}^{(1)}\nonumber\\
&&
+\frac{p_\mu}{2m}\left(\left[\overline{\Sigma_S^{\alpha\nu}(p_\alpha A_\nu)}^{(0)}\right]_{\text{PB}}+\left[\overline{\Sigma_P(p^\nu V_\nu)}^{(0)}\right]_{\text{PB}}\right)-\frac{m}{2}\left(\left[\overline{\Sigma_{S\mu\nu}A^\nu}^{(0)}\right]_{\text{PB}}+\left[\overline{\Sigma_PV_\mu}^{(0)}\right]_{\text{PB}}\right)\nonumber\\
&&
+\frac{1}{2}\epsilon_{\nu\mu\alpha\beta}\left[\overline{\Sigma_A^\nu(p^\alpha A^\beta)}^{(0)}\right]_{\text{PB}}
-\frac{1}{2}\epsilon_{\mu\nu\alpha\beta}\left(\overline{(\partial^\beta\Sigma_V^\nu)V^\alpha}^{(0)}-\overline{(\partial^\beta\Sigma_A^\nu)A^\alpha}^{(0)}\right)\nonumber\\
&&
-\frac{1}{2m}\left(p_\alpha\partial^\alpha(\overline{\Sigma_{S\mu\nu}A^\nu}^{(0)})-p^\nu\partial^\alpha(\overline{\Sigma_{S\mu\nu}A_\alpha}^{(0)})
-p_\mu\partial^\alpha(\overline{\Sigma_{S\alpha\nu}A^\nu}^{(0)})+p^\nu\partial^\alpha(\overline{\Sigma_{S\alpha\nu}A_\mu}^{(0)})\right),
\end{eqnarray}
where the hat operator is defined as $\overline{AB}^{(1)}=\overline{A^{(1)}B^{(0)}}+\overline{A^{(0)}B^{(1)}}$. The first line of the collision terms comes from dynamical processes, with the same structure as the collision terms in classical limit (\ref{TVA0}). The last three lines in both transport equations are related to the derivatives of the self-energies and distribution functions, arising from the inhomogeneous medium and generating the coupling between the vector and axial-vector components. As we will see in the following, this inhomogeneous effect is essential for spin polarization from thermal vorticity. 

\section{Anderson-Witting RTA}
\label {s3}
Relaxation time approximation is a widely used simplification to the collision kernel of the Boltzmann equation, its key point is assuming a relaxation time scale for systems close to equilibrium. However, this single time scale has not been proved to be suitable also for a multi-component quantum transport theory such as the spin transport. In this section, we revisit the commonly adopted RTA for the fermion Wigner function.

As a most general form of the relaxation time, one usually takes $p^\mu\partial_\mu f=-u^\mu p_\mu/\tau\delta f$ for the distribution function $f$ with momentum-dependent relaxation time $\tau(p)$ and local velocity $u^\mu$ of the medium. This is first adopted by Anderson and Witting~\cite{Anderson:1974}. With this choice, the current and energy-momentum conservations are satisfied using the Landau-Lifschitz matching conditions. Considering the spin degree of freedom, the RTA for fermion Wigner function can be written as~\cite{Hakim:1992}
\begin{eqnarray}
\label{RTA0}
\left(\gamma^\mu p_\mu-m\right)W
+\frac{i\hbar}{2}\gamma^\mu\partial_\mu W
~=~-\frac{i\hbar}{2}\gamma^\mu u_\mu\frac{\delta W}{\tau},
\end{eqnarray}
where $\delta W(x,p)=W(x,p)-W_{eq}(x,p)$ is the fluctuation of the Wigner function near equilibrium, and the four fluid velocity of the medium $u^\mu$ can be determined by the Landau matching condition $u_\mu(J^\mu-J_{eq}^\mu)=0$ and $u_\mu(T^{\mu\nu}-T_{eq}^{\mu\nu})=0$. The factor $\gamma^\mu u_\mu$ in (\ref{RTA0}) guarantees that the transport equation for any spin component of $W$ has the correct relaxation time structure such as $-u^\mu p_\mu/\tau\delta f$. Taking the spin decomposition of the kinetic equation (\ref{RTA0}), one obtains ten equations in A-W RTA,
\begin{eqnarray}
\label{compoAW1}
p_\mu V^\mu-m F &=& 0,\nonumber\\
2mP+ \hbar\partial_\mu A^\mu &=& -\frac{\hbar}{\tau} u_\mu\,\delta A^\mu,\nonumber\\
p_\mu F-mV_\mu+\frac{\hbar}{2}\partial^\nu{S}_{\mu\nu} &=& -\frac{\hbar}{2\tau}u^\nu\delta S_{\mu\nu},\nonumber\\
-\frac{1}{2}\epsilon_{\mu\nu\alpha\beta} p^\nu S^{\alpha\beta}-m A_\mu+\frac{\hbar}{2}\partial_\mu P &=& -\frac{\hbar}{2\tau}u_\mu \delta P,\nonumber\\
-2\epsilon_{\mu\nu\alpha\beta}p^\alpha A^\beta -2m S_{\mu\nu} +\hbar\partial_{[\mu} V_{\nu]} &=& -\frac{\hbar}{\tau}u_{[\mu} \delta V_{\nu]}
\end{eqnarray}
and 
\begin{eqnarray}
\label{compoAW2}
\hbar\partial_\mu V^\mu &=& -\frac{\hbar}{\tau} u_\mu\,\delta V^\mu,\nonumber\\
p_\mu A^\mu &=& 0,\nonumber\\
- p^\nu S_{\mu\nu}+\frac{\hbar}{2}\partial_\mu F &=& -\frac{\hbar}{2\tau}u_\mu \delta F,\nonumber\\
2p_\mu P+\frac{\hbar}{2}\epsilon_{\mu\nu\alpha\beta}\partial^\nu S^{\alpha\beta} &=& -\frac{\hbar}{2\tau}\epsilon_{\mu\nu\alpha\beta}u^\nu\delta S^{\alpha\beta},\nonumber\\
p_{[\mu} V_{\nu]} +\frac{\hbar}{2}\epsilon_{\mu\nu\alpha\beta}\partial^\alpha A^\beta &=& -\frac{\hbar}{2\tau}\epsilon_{\mu\nu\alpha\beta}u^\alpha\delta A^\beta.
\end{eqnarray}

The conservations of current,  energy momentum tensor and angular momentum tensor can be verified from the above equations. The ensemble average of the canonical current and energy momentum tensor are related to the component of the Wigner function through the moments of $V_\mu$, $\langle J_\mu\rangle=\int d^4 p V_\mu$ and $\langle T_{\mu\nu}\rangle=\int d^4 p p_\nu V_\mu$. Meanwhile, the ensemble average of the spin tensor can be expressed in terms of the Wigner function by $\langle S^{\lambda,\mu\nu}\rangle=-\hbar/2\int d^4 p\epsilon^{\lambda\mu\nu\alpha}A_\alpha$. The conservation laws can be verified through the kinetic equations together with the Landau matching conditions. Taking four momentum integral of the first equation in (\ref{compoAW2}) gives the current conservation, multiplying the equation by $p^\nu$ and then integrating it leads to the energy and momentum conservation, and the angular momentum conservation comes from the integration of the last equation in (\ref{compoAW2}),
\begin{eqnarray}
\label{conservation}
\partial_\mu J^\mu&=&-\frac{1}{\tau} u_\mu\delta J^\mu,\nonumber\\
\partial_\mu  T^{\mu\nu}&=&-\frac{1}{\tau} u_\mu \delta T^{\mu\nu},\nonumber\\
\partial_\alpha J^{\alpha,\mu\nu}&=&
T^{[\mu\nu]}+\partial_\alpha S^{\alpha,\mu\nu}=-\frac{1}{\tau}u_\alpha\delta S^{\alpha,\mu\nu}.
\end{eqnarray}
The current and energy-momentum conservations are guaranteed by applying the Landau matching conditions $u_\mu\delta J^{\mu}=0$ and $u_\mu \delta T^{\mu\nu}=0$. As the total angular momentum should be a conserved quantity, there could be another matching condition corresponding to the total angular momentum $ u_\lambda\delta J^{\lambda,\mu\nu}=0$. Using the definition of $J^{\lambda,\mu\nu}$, its contraction with $u_\lambda$ is given by $u_\lambda\delta J^{\lambda,\mu\nu}=u_\lambda[x^\mu \delta T^{\lambda\nu}-x^\nu \delta T^{\lambda\mu}+\delta S^{\lambda,\mu\nu}]=u_\lambda\delta S^{\lambda,\mu\nu}=0$. With this additional matching condition, the last equation in (\ref{conservation}) means total angular momentum conservation.

The RTA can be viewed as a rough recapitulation of the complicated microscopic mechanism that brings the system towards equilibrium, while the detailed collision terms contain much more information than the RTA. Here we list some of the difference in spin transport by directly comparing the A-W RTA (\ref{RTA0}) with the Kadanoff-Baym equation (\ref{KB}). With the A-W RTA (\ref{RTA0}), one has always $p_\mu V^\mu-m F=0$ and $p_\mu A^\mu=0$ despite of the existence of the relaxation process, while from the collision terms, one has $p_\mu V^\mu-m F=i\hbar/4C_F$ and $2p_\mu A^\mu=\hbar/2D_P$ with collision terms $C_F$ and $D_P$ which are in general nonzero. Secondly, at the classical level and first order in $\hbar$, the A-W RTA (\ref{RTA0}) indicates that the spin components are all on the mass shell $(p^2-m^2)V_\mu^{(0),(1)}=0$ and $(p^2-m^2)A_\mu^{(0),(1)}=0$, while from the Kadanoff-Baym equation, the on-shell condition is modified in quantum case. 

Without taking semi-classical expansion, the transport equations of the scalar and axial-vector components can be derived from (\ref{compoAW2}). Taking into account the asymmetric property of the tensor component $S_{\mu\nu}$ which results in $p^\mu p^\nu S_{\mu\nu}=0$, the scalar component $F$ experiences only damping effect when relaxing towards equilibrium,
\begin{equation}
\label{tr_S_AW}
p^\mu\partial_\mu F = -{u^\mu p_\mu\over \tau}\delta F.
\end{equation}
On the other hand, the transport equation for the axial-vector component contains both damping and polarization effect,
\begin{equation}
\label{tr_A_AW}
p^\nu \partial_\nu A_\mu = -{u^\nu p_\nu\over \tau}\delta A_\mu -\frac{\hbar}{2m}\epsilon_{\mu\nu\alpha\beta}p^\beta\left(\partial^\alpha \frac{u^\nu}{\tau}\right)\delta F.
\end{equation}
The damping rate for the axial-vector component is the same as that for the scalar component. This, as we will see in the following, is the same as what one obtains from the Kadanoff-Baym equation. The second term on the right-hand side of (\ref{tr_A_AW}) indicates a spin polarization effect arising from the coupling with the off-equilibrium fluctuation of the scalar component. There is a factor of $\hbar$ in the second term, it indicates that the polarization effect is a quantum effect. Note that, the fluctuation of the axial-vector component is at least at the first order of $\hbar$, while the fluctuation of the scalar component can be at the classical level.   

The temperature dependence of the relaxation time can be taken as $\tau=5\bar \eta/T$~\cite{Denicol:2011fa, Jaiswal:2013vta} with the shear viscosity to entropy density ratio $\bar{\eta}=\eta/s$. Considering the classical solution of the scalar component $F^{(0)}(x,p)=2m\delta(p^2-m^2)f_V(x,p)=m/E_p\left[\delta(p_0-E_p)f(x,{\bf p})+\delta(p_0+E_p)\bar f(x,-{\bf p})\right]$ with the fermion number distributions $f$ and $\bar f$, the transport equation for the axial-vector component can be rewritten as 
\begin{equation}
\label{tr_A_AW_1}
p^\nu\partial_\nu A_\mu = -{u^\nu p_\nu\over \tau}\delta A_\mu +{1\over \tau}\frac{\hbar}{T}p^\nu \widetilde{\omega}^{T}_{\mu\nu}\delta f_V, 
\end{equation}
where $\omega^{T}_{\mu\nu}$ is the $T$-vorticity defined as $\omega^{T}_{\mu\nu}=1/2[\partial_\mu(T u_\nu)-\partial_\nu(T u_\mu)]$, and $\widetilde{\omega}^{T}_{\mu\nu}$ is the dual $T$-vorticity defined as $\widetilde{\omega}^{T}_{\mu\nu}=1/2\epsilon_{\mu\nu\rho\sigma} \omega_T^{\rho\sigma}$. Eq.(\ref{tr_A_AW_1}) indicates that non-equilibrium spin polarization is generated by the $T$-vorticity, which can give the correct quadrupole distribution in local-polarization~\cite{Wu:2019eyi} and may give a hint for solving the "sign problem" in local-polarization.

To clearly illustrate the polarization timescale given by A-W RTA, we consider a homogeneous temperature. In this case $\widetilde{\omega}^{T}_{\mu\nu}/T$ in the second term of (\ref{tr_A_AW_1}) is reduced to the dual kinetic vorticity $\widetilde{\omega}_{\mu\nu}^{K}=1/2\epsilon_{\mu\nu\rho\sigma}\partial^\rho u^\sigma$ which can be decomposed into the vorticity $\omega_\mu=1/2\epsilon_{\mu\nu\rho\sigma}u^\nu\partial^\rho u^\sigma$ and acceleration $\varepsilon_\mu=1/2u_\lambda \partial^\lambda u_\mu$ as $\widetilde{\omega}^{K}_{\mu\nu}=\omega_\mu u_\nu-\omega_\nu u_\mu+\epsilon_{\mu\nu\rho\sigma}\varepsilon^\rho u^\sigma$. For a system with static vorticity without acceleration, $\widetilde{\omega}^{K}_{\mu\nu}=\omega_\mu u_\nu-\omega_\nu u_\mu$, the transport equation for the spatial component ${\bf A}$ which means the spin density becomes
\begin{equation}
\label{tr_A_AW_2}
p^\mu\partial_\mu {\bf A} = -{E_p\over \tau}\delta {\bf A} +{E_p\over \tau}\left(\hbar{\bm \omega}\delta f_V\right)
\end{equation}
in the local rest frame with $u_\mu=(1,{\bf 0})$. Since the spin distribution and its fluctuation around equilibrium are at least at the first order in $\hbar$ and the number fluctuation can be at the zeroth order, the both sides of (\ref{tr_A_AW_2}) are at the same order. The factor $\hbar{\bm \omega} \delta f_V$ has the same structure as the axial-vector component and can be explained as a non-equilibrium spin fluctuation induced by vorticity and number fluctuation. The timescale for polarization is the same as for the damping. There are both damping and polarization effects in the direction parallel to the vorticity and only damping in directions perpendicular to the vorticity. In conclusion, the A-W RTA (\ref{RTA0}) can describe the spin polarization with the same time scale for polarization and damping. 

\section{RTA from Kadanoff-Baym equation}
\label {s4}
The general transport equations derived from the Kadanoff-Baym equations (\ref{KB}) contain all possible collision information. The full collision terms could be taken as a starting point to study the relaxation process in a system close to equilibrium. Any distribution function of the system can be separated into an equilibrium part and a fluctuation part. The equilibrium distribution eliminates the collision kernel, where the gain term and loss term cancel with each other. The fluctuation part satisfies the transport equation, with the collision kernel linearly proportional to the fluctuation. 
 
We first consider the transport equations at classical level. Using the relation $F^{(0)}=p^\mu/m V_\mu^{(0)}$, the coupled transport equations (\ref{TVA0}) for $V_\mu^{(0)}$ and $A_\mu^{(0)}$ become the coupled equations for the number density and spin density $F^{(0)}$ and $A_\mu^{(0)}$, 
\begin{eqnarray}
\label{0th_transport}
p^\nu\partial_\nu F^{(0)} &=& m\overline{\Sigma_F F}^{(0)} +p_\nu\overline{\Sigma_V^\nu F}^{(0)} +\frac{1}{2}\epsilon_{\alpha\beta\lambda\nu}p^\nu\overline{\Sigma_S^{\alpha\beta}A^\lambda}^{(0)}-m\overline{\Sigma_A^\nu A_\nu}^{(0)},\nonumber\\
p^\nu\partial_\nu A^{(0)}_\mu &=& m\overline{\Sigma_FA_\mu}^{(0)} +p_\nu\overline{\Sigma_V^\nu A_\mu}^{(0)} +\frac{1}{2}\epsilon_{\alpha\beta\lambda\mu}p^\lambda \overline{\Sigma_S^{\alpha\beta}F}^{(0)} +m\overline{\Sigma_{A\mu}F}^{(0)}-\frac{p_\mu p_\nu }{m}\overline{\Sigma_A^\nu F}^{(0)}.
\end{eqnarray}
When the system is close to equilibrium, the self-energies in the transport equations are calculated in equilibrium state. Considering the detailed balance for the equilibrium state, the above transport equations can be grouped into a compact form, 
\begin{eqnarray}
\label{0th_delta_transport}
p^\nu\partial_\nu\left(\begin{array}{c}
F^{(0)}\\
A^{(0)}_\mu\\
\end{array}\right)=\left(\begin{array}{cc}
\omega_0^{(0)} & \omega^{(0)\mu}_s\\
-\omega_{s\mu}^{(0)} & \omega_0^{(0)}\\
\end{array}\right)\left(\begin{array}{c}
\delta F^{(0)}\\
\delta A^{(0)}_\mu\\
\end{array}\right),
\end{eqnarray}
where the diagonal element $\omega_0$ is the dissipation rate, and the off-diagonal element $\omega_{s\mu}$ is the interaction rate of the exchange between axial-charge and charge, namely the exchange between orbital and spin angular momentum,  
\begin{eqnarray}
\omega_0^{(0)}(p) &=& m\widehat\Sigma_F^{(0)} +p^\nu\widehat\Sigma_{V\nu}^{(0)},\nonumber\\
\omega_{s\mu}^{(0)}(p) &=& \frac{1}{2}\epsilon_{\alpha\beta\mu\nu}p^\nu\widehat{\Sigma}_{S}^{(0)\alpha\beta}-m\widehat\Sigma_{A\mu}^{(0)}+\frac{p_\mu p^\nu}{m}\widehat\Sigma_{A\nu}^{(0)}
\end{eqnarray}  
with the definition of $\widehat\Sigma=\overline\Sigma+\Sigma$. In deriving (\ref{0th_delta_transport}) from (\ref{0th_transport}), we have used $\delta\overline F^{(0)}=-\delta F^{(0)}$ and $\delta\overline A^{(0)}_\mu=-\delta A^{(0)}_\mu$. These two relations follow from $\bar{f}_V^{(0)}=1-{f}_V^{(0)}$ and $\bar{f}_A^{(0)}=-{f}_A^{(0)}$ which are valid for both non-equilibrium and equilibrium distributions. Since $A^{(0)}_\mu$ is perpendicular to $p_\mu$, $p^\mu A^{(0)}_\mu=0$, the fluctuation should be perpendicular to $p_\mu$ too, $p^\mu\delta A^{(0)}_\mu=0$, which leads to $1/mp^\mu p_\nu\widehat{\Sigma_A^{(0)\nu}}\delta A^{(0)}_\mu=0$ and in turn the symmetric representation of the matrix equation (\ref{0th_delta_transport}). 

Since the fluctuation is around the equilibrium, the damping and polarization rates in transport equations (\ref{0th_delta_transport}) are calculated with the equilibrium distributions. Projecting the transport equation for $A^{(0)}_\mu$ to $p^\mu$, the constraints $p^\mu A^{(0)}_\mu=0$ and $p^\mu\delta A_\mu^{(0)}=0$ lead to the restriction $p^\mu\omega_{s\mu}=0$. This means that in general case the polarization vector $\omega_{s\mu}$ has only three independent Lorentz components. Considering the fact that the elimination of collision terms at classical level requires $A^{(0)}_\mu=0$~\cite{Wang:2020pej}, the axial and tensor components of the self-energy would vanish by power counting, ${\Sigma}_A^{(0)\mu}={\Sigma}_S^{(0)\mu\nu}=0$. Therefore, the polarization rate disappears at classical level, $\omega_{s\mu}^{(0)}=0$, which means no polarization effect at classical level, as one would expect. In this case, there is only damping effect, the transport equation becomes diagonal, 
 \begin{eqnarray}
p^\nu\partial_\nu\left(\begin{array}{c}
F^{(0)}\\
A^{(0)}_\mu\\
\end{array}\right)=\omega_0^{(0)}\left(\begin{array}{c}
\delta F^{(0)}\\
\delta A^{(0)}_\mu\\
\end{array}\right).
\end{eqnarray}

The spin polarization can be generated from the spin interaction with electromagnetic and vortical fields, which all happen at order $\mathcal{O}(\hbar)$. Without considering the background electromagnetic field, taking into account the classical self-energies $\Sigma_A^{(0)\mu}=\Sigma_{S}^{(0)\mu\nu}=0$, and using the first-order constraints $F^{(1)}=p^\mu/mV^{(1)}_\mu-i/(4m)C_F^{(0)}=p^\mu/mV^{(1)}_\mu$ ($C_F^{(0)}=0$, see Appendix \ref{a1}), the coupled transport equations (\ref{TVA1}) for $V_\mu^{(1)}$ and $A_\mu^{(1)}$ are changed to be the coupled equations for $F^{(1)}$ and $A_\mu^{(1)}$,  
\begin{eqnarray}
\label{TSA1}
p^\nu\partial_\nu F^{(1)} &=& m\overline{\Sigma_F F}^{(1)}+p_\nu\overline{\Sigma_V^\nu F}^{(1)}
-m\overline{\Sigma_{A\nu}^{(1)}A^{(0)\nu}}+\frac{p^\alpha p^\nu}{m}\overline{\Sigma_{A\nu}^{(1)}A^{(0)}_\alpha}\nonumber\\
&& +\frac{1}{2}\epsilon_{\alpha\nu\sigma\rho}p^\nu\overline{\Sigma_{S}^{(1)\sigma\rho}A^{(0)\alpha}}
+\frac{1}{2m}\epsilon_{\alpha\nu\rho\sigma}p^\rho\overline{(\partial^\sigma\Sigma_{V}^{\nu})A^{\alpha}}^{(0)}
-\frac{1}{2}\overline{(\partial_\mu\Sigma_P)A^{\mu}}^{(0)},\nonumber\\
p^\nu\partial_\nu A_\mu^{(1)} &=& m\overline{\Sigma_FA_{\mu}}^{(1)}+p^\nu\overline{\Sigma_{V\nu}A_{\mu}}^{(1)}
+m\overline{\Sigma_{A\mu}^{(1)}F^{(0)}}-\frac{p_\mu p^\nu}{m}\overline{\Sigma_{A\nu}^{(1)}F^{(0)}}\nonumber\\
&&
-\frac{1}{2}\epsilon_{\mu\nu\sigma\rho}p^\nu\overline{\Sigma_S^{(1)\sigma\rho}F^{(0)}}
-\frac{1}{2m}\epsilon_{\mu\nu\rho\sigma}p^\rho\overline{(\partial^\sigma\Sigma_{V}^{\nu})F}^{(0)}
+\frac{1}{2}\overline{(\partial_\mu\Sigma_P)F}^{(0)}.
\end{eqnarray}
While the term $p_\mu p^\nu/m\overline{\Sigma_{A\nu}^{(1)}A^{(0)\mu}}$ in the transport equation for $F^{(1)}$ vanishes due to the condition $p_\mu A^{(0)\mu}=0$, it is kept to maintain the symmetry between the two equations. Adding the classical and first-order transport equations (\ref{0th_transport}) and (\ref{TSA1}), we have for $F=F^{(0)}+\hbar F^{(1)}$ and $A_\mu=A^{(0)}_\mu+\hbar A^{(1)}_\mu$,
\begin{eqnarray}
p^\nu\partial_\nu F &=& m\overline{\Sigma_F F}+p_\nu\overline{\Sigma_V^\nu F}\nonumber\\
&&
+\hbar\left[-m\overline{\Sigma_{A}^{\nu}A_\nu}+\frac{1}{2}\epsilon_{\alpha\nu\sigma\rho}p^\nu\overline{\Sigma_S^{\sigma\rho}A^\alpha}+\frac{p_\alpha p_\nu}{m}\overline{\Sigma_A^\nu A^\alpha}
+\frac{1}{2m}\epsilon_{\alpha\nu\rho\sigma}p^\rho\overline{(\partial^\sigma\Sigma_V^\nu)A^\alpha}
-\frac{1}{2}\overline{(\partial^\nu\Sigma_P)A_{\nu}}\right],\nonumber\\
p^\nu\partial_\nu A_\mu
&=&
m\overline{\Sigma_F A_{\mu}}+p_\nu\overline{\Sigma_V^\nu A_{\mu}}
\nonumber\\
&&
-\hbar\left[-m\overline{\Sigma_{A\mu} F}+\frac{1}{2}\epsilon_{\mu\nu\sigma\rho}p^\nu\overline{\Sigma_S^{\sigma\rho} F}
+\frac{p_\mu p^\nu}{m}\overline{\Sigma_{A\nu} F}+\frac{1}{2m}\epsilon_{\mu\nu\rho\sigma}p^\rho\overline{(\partial^\sigma\Sigma_{V}^{\nu}) F}
-\frac{1}{2}\overline{(\partial_\mu\Sigma_P) F}\right],
\end{eqnarray}
with $\Sigma=\Sigma^{(0)}+\hbar\Sigma^{(1)}$. For systems close to equilibrium, the equilibrium state eliminates the collision terms as required by the detailed balance principle, and the collision terms are linearly proportional to the non-equilibrium fluctuations $\delta F$ and $\delta A^\mu$, 
\begin{eqnarray}
\label{KB_RTA}
p^\nu\partial_\nu\left(\begin{array}{c}
F\\
A_\mu\\
\end{array}\right)=\left(\begin{array}{cc}
\omega_0 & \hbar\omega^\mu_s\\
-\hbar\omega_{s\mu} & \omega_0\\
\end{array}\right)\left(\begin{array}{c}
 \delta F\\
\delta A_\mu\\
\end{array}\right)
\end{eqnarray}
with the interaction rates
\begin{eqnarray}
\label{interactionrate}
\omega_0(p) &=& m\widehat\Sigma_F +p_\nu\widehat\Sigma_V^\nu\nonumber\\
\omega_{s\mu}(p) &=& -m\widehat\Sigma_{A\mu}+\frac{1}{2}\epsilon_{\mu\nu\sigma\rho}p^\nu\widehat\Sigma_S^{\sigma\rho}+\frac{p_\mu p_\nu}{m}\widehat\Sigma_A^\nu
+\frac{1}{2m}\epsilon_{\mu\nu\rho\sigma}p^\rho(\partial^\sigma\widehat\Sigma_{V}^{\nu})-\frac{1}{2}\partial_\mu\widehat\Sigma_P.
\end{eqnarray}
For a specific interaction, the damping rate $\omega_0$ and polarization rate $\omega_{s\mu}$ and in turn the relaxation times $\tau_0$ and $\tau_{s\mu}$ can be explicitly calculated.

The transport equations (\ref{tr_S_AW}) and (\ref{tr_A_AW}) in A-W RTA and the equations (\ref{KB_RTA}) in Kadanoff-Baym formalism share the property that the damping rates for number density and spin density are the same. While the two fluctuations $\delta F$ and $\delta A_\mu$ are coupled to each other via the transport equations (\ref{KB_RTA}), the coupling appeared in the equation for $\delta F$ is $\sim \hbar^2$, due to the fact that $\delta A_\mu$ itself is at order $\mathcal{O}(\hbar)$, and then can safely be neglected. The transport equations (\ref{KB_RTA}) and (\ref{interactionrate}) provide a dynamical way to microscopically calculate the relaxation times for different spin components. For instance, the relaxation time corresponding to the damping process in the local rest frame is 
\begin{equation}
\tau_0 = -{E_p\over \omega_0}.
\end{equation}

Considering the polarization effect, from the A-W RTA (\ref{tr_A_AW_1}), spin could be polarized by the T-vorticity, and the timescale is the same as for the damping. Namely, charge and spin will reach  thermalization at the same time, as required by the assumption of a single timescale. However, considering the interaction rates (\ref{interactionrate}) in Kadanoff-Baym formalism, spin is polarized by the thermal-vorticity instead of $T$-vorticity, and the timescales for charge thermalization and spin thermalization are in general different. In the following, we will take a specific model as an example to calculate the interaction rates.  

\section{Interaction rates in NJL model}
\label{s5}
In this section, we adopt a specific interaction to calculate the spin-dependent interaction rates in (\ref{interactionrate}). To see clearly the difference in timescales of damping and polarization and to avoid the model dependence as much as possible, we focus on the ratio of the two rates $\omega_{s\mu}$ and $\omega_0$. We take a two-flavor Nambu-Jona-Lasinio model (NJL) with only scalar interaction, 
\begin{equation}
\mathcal{L}=\bar\psi(i\hbar\gamma^\mu \partial_\mu-m)\psi+G(\bar\psi\psi)^2,
\end{equation}
where $G$ is the coupling constant of the contact interaction with dimension of (GeV)$^{-2}$. To make semi-classical expansion of the obtained kinetic equations, we have explicitly shown the $\hbar$ dependence. Considering the fermionic two-by-two scattering to the first order of $1/N_c$ expansion, where $N_c$ is the number of colors, the quark self-energy can be written as
\begin{equation}
\label{selfenergy1}
\Sigma^{>}(p) = G^2\int dP S^{>}(p_1)\text{Tr}\left[S^{<}(p_2)S^{>}(p_3)\right]
\end{equation}
with the momentum integral $\int dP=\int d^4p_1 d^4p_2 d^4p_3/(2\pi)^8\delta(p-p_1+p_2-p_3)$. At classical level and order ${\mathcal O}(\hbar)$ the integrated function can be expanded as  
\begin{eqnarray}
\label{01}
\left\{S^{>}(p_1)\text{Tr}\left[S^{<}(p_2)S^{>}(p_3)\right]\right\}^{(0)} &=& S^{>(0)}(p_1)\text{Tr}\left[S^{<(0)}(p_2)S^{>(0)}(p_3)\right],\nonumber\\
\left\{S^{>}(p_1)\text{Tr}\left[S^{<}(p_2)S^{>}(p_3)\right]\right\}^{(1)} &=& S^{>(1)}(p_1)\text{Tr}\left[S^{<(0)}(p_2)S^{>(0)}(p_3)\right]+S^{>(0)}(p_1)\text{Tr}\left[S^{<(1)}(p_2)S^{>(0)}(p_3)\right]\nonumber\\
&& +S^{>(0)}(p_1)\text{Tr}\left[S^{<(0)}(p_2)S^{>(1)}(p_3)\right].
\end{eqnarray}

For systems close to equilibrium, the interaction rates are calculated in equilibrium state. In this case, it is analytically demonstrated that~\cite{Wang:2020pej} the non-zero equilibrium Wigner function contains only the scalar and vector components $F^{(0)}$ and $V_\mu^{(0)}$ at classical level and axial-vector and tensor components $A_\mu^{(1)}$ and $S_{\mu\nu}^{(1)}$ at order ${\mathcal O}(\hbar)$. Therefore, the trace in (\ref{01}) becomes simple, 
\begin{eqnarray}
\label{trace}
&& \text{Tr}\left[S^{<(0)}(p_2)S^{>(0)}(p_3)\right] = \frac{m^2+p_2\cdot p_3}{m^2}F^{(0)}(p_2)\overline F^{(0)}(p_3),\nonumber\\
&& \text{Tr}\left[S^{<(1)}(p_2)S^{>(0)}(p_3)\right] = \text{Tr}\left[S^{<(0)}(p_2)S^{>(1)}(p_3)\right]=0.
\end{eqnarray} 
Substituting the result into the self-energy defined in (\ref{selfenergy1}), its non-zero components include only $\Sigma_F$ and $\Sigma_V^\mu$ at classical level and $\Sigma_A^\mu$ and $\Sigma_S^{\mu\nu}$ at order $\mathcal{O}(\hbar)$. The damping and polarization rates defined by (\ref{interactionrate}) can be evaluated explicitly in terms of the equilibrium components $F$ and $A_\mu$, 
\begin{eqnarray}
\label{omega}
\omega_0(p) &=& G^2\int dP{m^2+p_2\cdot p_3\over m^3}(m^2+p\cdot p_1)\left(\overline F^{(0)}(p_1) F^{(0)}(p_2)\overline F^{(0)}(p_3)+F^{(0)}(p_1)\overline F^{(0)}(p_2) F^{(0)}(p_3)\right),\nonumber\\
\omega_{s\mu}(p) &=& G^2\int dP {m^2+p_2\cdot p_3\over m^4}\Big[-m(m^2+p\cdot p_1)\left(\overline A_\mu^{(1)}(p_1) F^{(0)}(p_2)\overline F^{(0)}(p_3)+A_\mu^{(1)}(p_1)\overline F^{(0)}(p_2)F^{(0)}(p_3)\right)\nonumber\\
&&\qquad\qquad\qquad\qquad\quad+m(p_\mu+p_{1\mu})p^\nu\left(\overline A_\nu^{(1)}(p_1) F^{(0)}(p_2)\overline F^{(0)}(p_3)+A_\nu^{(1)}(p_1)\overline F^{(0)}(p_2)F^{(0)}(p_3)\right)\nonumber\\
&&\qquad\qquad\qquad\qquad\quad+\frac{1}{2}\epsilon_{\mu\nu\alpha\beta}p^\nu p_1^\beta\left([\partial^\alpha\overline F^{(0)}(p_1)]F^{(0)}(p_2)\overline F^{(0)}(p_3)+ [\partial^\alpha F^{(0)}(p_1)]\overline F^{(0)}(p_2)F^{(0)}(p_3)\right)\nonumber\\
&&\qquad\qquad\qquad\qquad\quad+\frac{1}{2}\epsilon_{\mu\nu\alpha\beta}p^\nu p_1^\beta\partial^\alpha\left(\overline F^{(0)}(p_1)F^{(0)}(p_2)\overline F^{(0)}(p_3)+ F^{(0)}(p_1)\overline F^{(0)}(p_2)F^{(0)}(p_3)\right)\Big].
\end{eqnarray}
Note that, the equilibrium distributions $F$ and $A_\mu$ depend on space-time through the local velocity $u_\mu(x)$ and temperature $T(x)$ of the medium, and can be expressed in terms of the Fermi-Dirac distributions $f$ and $\bar f$, 
\begin{eqnarray}
\label{equilibrium}
F^{(0)}(x,p) &=& \frac{m}{E_p}\left[\delta(p_0-E_p)f(x,{\bf p})+\delta(p_0+E_p)\bar f(x,-{\bf p})\right],\nonumber\\
A^{(1)}_\mu(x,p) &=& -\tilde \omega_{\mu\nu}^{th}p^\nu\frac{1}{2E_p}\left[\delta(p_0-E_p)f'(x,{\bf p})+\delta(p_0+E_p)\bar f'(x,-{\bf p})\right],
\end{eqnarray}
where $\omega_{\sigma\lambda}^{th}=1/2(\partial_\sigma\beta_\lambda-\partial_\lambda\beta_\sigma)$ is the thermal vorticity with $\beta_\lambda=u_\lambda/T$, $\tilde\omega_{\mu\nu}^{th}=1/2\epsilon_{\mu\nu\sigma\lambda}\omega^{\sigma\lambda}_{th}=1/2\epsilon_{\mu\nu\sigma\lambda}\partial^\sigma\beta^\lambda$ is the dual thermal vorticity, and $f'$ and $\bar f'$ are the derivative of the distributions $f'=\partial f/\partial (p^\mu\beta_\mu)$ and $\bar f' =\partial \bar f/\partial (p^\mu \beta_\mu)$. Note that the equilibrium axial-vector distribution which eliminates the collision terms is the global equilibrium one~\cite{Wang:2020pej}, the spin should be polarized by the thermal vorticity. 

Let's first prove the proportional relation between the spin polarization rate $\omega_{s\mu}$ and the thermal vorticity $\omega_{\mu\nu}^{th}$ in the Kadanoff-Baym formalism, which is in contrast with the A-W RTA (\ref{tr_A_AW_1}) where the polarization is generated by the $T$-vorticity. Any of the terms in the first two lines of $\omega_{s\mu}$ contains the axial-vector component which is linear in the thermal vorticity, see (\ref{equilibrium}). For the other terms, any of them includes the coordinate derivative of the scalar component. Let's consider as an example the term $p_2^\rho\epsilon_{\mu\nu\alpha\beta}p^\nu p_1^\beta\partial^\alpha\beta_\rho f'(p_2)$. Employing the Schouten identity and Killing condition $\partial_\mu\beta_\nu+\partial_\nu\beta_\mu=0$ in global equilibrium state~\cite{Becattini:2016stj}, one obtains
\begin{equation}
p_2^\rho\epsilon_{\mu\nu\alpha\beta}p^\nu p_1^\beta\partial^\alpha\beta_\rho f'(p_2) = \left[p_{2\mu}p^\nu p_1^\rho \tilde \omega_{\nu\rho}^{th}
-(p\cdot p_{2}) p_1^\nu \tilde \omega_{\mu\nu}^{th}+(p_1\cdot p_{2})p^\nu \tilde \omega_{\mu\nu}^{th}\right] f'(p_2),
\end{equation}
other terms related to derivatives also have the same structure. Therefore, the spin polarization is indeed generated by the thermal vorticity, instead of the $T$-vorticity. The Killing condition is used here, because for the axial-vector component the collision terms vanish only in global equilibrium state where the Killing condition exists. We should emphasize that, the conclusion of spin polarization by $T$-vorticity in the A-W RTA comes from the assumption of a single timescale $\tau$ and the usage of the relation $\tau=5\bar{\eta}/T$. In general, collision terms could generate all kinds of vorticity structures in off-equilibrium systems. However, for systems close to equilibrium, only thermal vorticity contributes. 

We now consider the timescale for polarization and the comparison with the timescale for damping in the Kadanoff-Baym formalism. To see clearly the difference between the two timescales and to avoid as much as possible the model dependence, we calculate the ratio between the two relaxation times $\tau_0$ and $\tau_{s\mu}$. For simplicity, we consider an equilibrium system with constant temperature. In this case, the dual thermal vorticity is reduced to $1/(2T)\epsilon_{\mu\nu\rho\sigma}\partial^\rho u^\sigma=1/T(\omega^\mu u^\nu-\omega^\nu u^\mu)$. If the global angular velocity is taken to be along the $z$-axis, ${\bm \omega}=\omega \hat{\bm z}$, the non-vanishing components of the dual thermal vorticity are only $\tilde \omega_{03}^{th}=-\tilde \omega_{30}^{th}=\omega/T$, and the spin polarization rate $\omega_{s\mu}$ in (\ref{omega}) is proportional to $\omega/T$. To compare with the A-W RTA (\ref{tr_A_AW_2}), we rewrite the transport equation for the axial-vector component in (\ref{KB_RTA}) as  
\begin{equation}
p^\mu\partial_\mu {\bf A} = -\Gamma_0\delta{\bf A} +{\bf \Gamma}_s \left(\hbar\omega\delta f_V\right),
\end{equation}
and calculate the ratio between the two interaction rates,
\begin{equation}
{\Gamma_{si}\over \Gamma_0} = {\tau_0\over \tau_{si}} = {2m\over \omega}{\omega_{si}\over \omega_0}.
\end{equation}

The damping rate $\omega_0$ contains contributions from three scattering channels allowed by energy-momentum conservation. Each term is accompanied by a product of three distribution functions $f$ and $\bar f$, corresponding to a scattering process of incoming quark and anti-quark with $f$ and outgoing quark and anti-quark with $\bar f$. To further simplify the numerical calculation, we consider in the following only quark-quark scatterings. In this case the damping rate becomes 
\begin{equation}
\label{omega0}
\omega_0(p) = 8G^2\int d{\bf P}\left(m^2+E_2 E_3-{\bf p}_2\cdot{\bf p}_3\right)\left(m^2+E E_1-{\bf p}\cdot{\bf p}_1\right)\left(\bar f_3f_2\bar f_1+f_3\bar f_2f_1\right)
\end{equation}
with the three-dimensional momentum integral $d{\bf P} = d^3{\bf p}_1 d^3{\bf p}_2 d^3{\bf p}_3/(8E_1E_2E_3(2\pi)^8)\delta({\bf p}-{\bf p}_1+{\bf p}_2-{\bf p}_3)$ and the definition of $f_i=f({\bf p}_i)$. To calculate the polarization rate $\omega_{s\mu}$, we substitute the equilibrium solutions (\ref{equilibrium}) into (\ref{omega}). There are still three channels allowed by energy-momentum conservation. One corresponds to the quark-quark scattering, and the other two are for the quark-antiquark scatterings. We consider again only the quark-quark scattering. In global equilibrium state, the quantum spin distribution $A^{(1)}_\mu$ has only one independent Lorentz component, $A^{(1)}_1(p)=A^{(1)}_2(p)=0, A^{(1)}_3(p)\neq 0$ and $A^{(1)}_0(p)=p_3/p_0 A^{(1)}_3(p)$. We calculate separately the polarization rate in the directions perpendicular and parallel to the rotation vector ${\bm \omega}=\omega\hat{\bf z}$. Taking $\mu=1$, the perpendicular polarization along the $x$-axis is 
\begin{eqnarray}
\label{omegasx}
\omega_{sx}(p) &=& 8G^2\frac{\omega}{2mT}\int d{\bf P}\left(m^2+E_2E_3-{\bf p}_2\cdot{\bf p}_3\right)(E_1p_z -Ep_{1z})\nonumber\\
&&\times\left[(p_x-p_{1x})\left(\bar f_2f_3-f_2\bar f_3\right)f'_1-p_{2x}\left(\bar f_1\bar f_3-f_1f_3\right)f'_2-p_{3x}\left(f_1\bar f_2-\bar f_1f_2\right)f'_3\right].
\end{eqnarray}
Different from the damping rate, the polarization rate contains the derivative of the distribution function, $f'$. Similarly, the parallel polarization $\omega_{sz}$ in the $z$-direction is evaluated as 
\begin{eqnarray}
\label{omegasz}
\omega_{sz}(p) &=& 8G^2\frac{\omega}{2mT}\int d{\bf P}\left(m^2+E_2E_3-{\bf p}_2\cdot {\bf p}_3\right)\nonumber\\
&&\times\big[\left(m^2(E_1+E)+E_1p_z^2-Ep_zp_{1z}\right)\left(\bar f_2f_3-f_2\bar f_3\right)f'_1+\left(E_1p_\perp\cdot p_{1\perp}-Ep_{1\perp}^2\right)\left(\bar f_2f_3-f_2\bar f_3\right)f'_1\nonumber\\
&&+\left(E_1p_\perp\cdot p_{2\perp}-Ep_{1\perp}\cdot p_{2\perp}\right)\left(\bar f_1\bar f_3-f_1f_3\right)f'_2+\left(E_1p_\perp\cdot  p_{3\perp}-Ep_{1\perp}\cdot p_{3\perp}\right)\left(f_1\bar f_2-\bar f_1f_2\right)f'_3\big].
\end{eqnarray}

We now calculate the ratio $\Gamma_{sz}/\Gamma_0$ as a function of the fireball temperature. Since we consider the quark interaction rates, the temperature is restricted to be larger than the critical temperature $T_c$ of the quark deconfinement phase transition. Due to the fact that the contact interaction in the NJL model is not renormalizable, we should use a momentum cutoff $\Lambda$ in the phase space integration to avoid divergence. The three parameters in the model, namely the quark mass $m$, coupling constant $G$ and momentum cutoff $\Lambda$ can be fixed by fitting the pion mass, pion decay constant and chiral condensate in vacuum~\cite{Klevansky:1992qe}. In this work, the momentum integral is carried out through Monte Carlo integration, with a hard cutoff $\Lambda=1.5$GeV, $G=0.775$GeV$^{-2}$, and bare mass $m=5$MeV, this gives the critical temperature of chiral phase transition $T_c\sim230$MeV. While either the damping or the polarization rate may strongly depend on the values of the parameters, the dependence of the ratio should be largely weakened. 

\begin{figure}[H]\centering
	\includegraphics[width=0.45\textwidth]{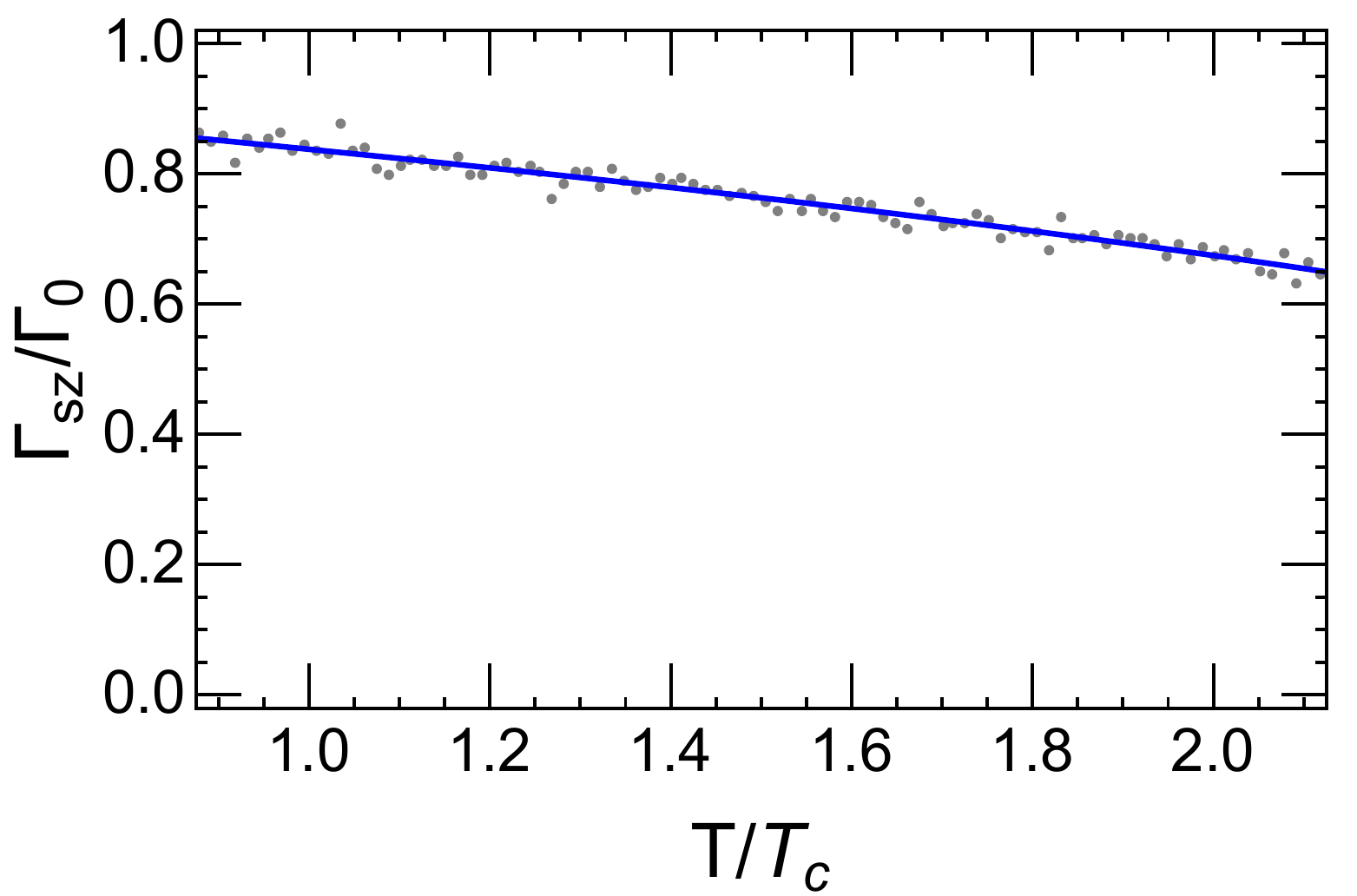}
	\caption{The temperature dependence of the ratio between polarization rate and damping rate $\Gamma_{sz}/\Gamma_0$ in the NJL model.}
	\label{fig1}
\end{figure}

The numerical result is shown in Fig.\ref{fig1}. While both the polarization rate and damping rate increase with temperature, their ratio decreases. This is qualitatively reasonable, as the system becomes more isotropic at high temperature. From the relation between interaction rate and relaxation time, $\Gamma_{si}/\Gamma_0=\tau_0/\tau_{si}$, the thermalization time scales for damping and polarization are different. In the considered temperature region $1<T/T_c<2$, the relaxation time for damping is smaller than that for polarization, $\tau_0/\tau_{sz}\sim 0.7$. This means that charge is thermalized earlier and spin is thermalized later. 

\section{Conclusion}
\label{s6}
The collision terms in quantum transport theory, which control the process from non-equilibrium to equilibrium of the system, can systematically be derived from the Kadanoff-Baym formalism order by order in $\hbar$, yet the calculation is complicated. The relaxation time approximation is an often used way to effectively describe the collision effect for systems close to equilibrium, at the expense of assuming a single time scale for different degrees of freedom. The RTA is successfully applied to the thermalization of particle number distribution. However, the assumption of a single time scale has not been proved to be suitable for a multi-component quantum transport theory such as the spin transport.  

We analyzed the property of collision terms when approaching to equilibrium with the Kadanoff-Baym equations, and made comparison with the commonly adopted Anderson-Witting RTA. We derived the classical and quantum transport equations for fermion number and spin distributions near equilibrium, and the collision terms are expressed in terms of the interaction rates determined at equilibrium. The time scales for number damping and spin damping are the same, this is the same result as obtained in A-W RTA. However, different from the study using A-W RTA where the spin polarization is induced by T-vorticity and the damping and polarization share the same relaxation time, the polarization in the Kadanoff-Baym formalism can only be generated by thermal vorticity and the two time scales for damping and polarization are different in general case. As a simple example, we explicitly calculated the temperature dependence of the ratio between the two relaxation times in the NJL model. The result shows that the number density is thermalized earlier than the spin density. To be realistic, the calculation will be extended to QED and QCD plasma in our future work. \\
\\
\noindent {\bf Acknowledgement}: The work is supported by Guangdong Major Project of Basic and Applied Basic Research No. 2020B0301030008 and NSFC grant Nos. 11890712, 12005112 and 12075129. ZW is also supported by the Postdoctoral Innovative Talent Support Program of Tsinghua University. 

\begin{appendix}
\section{Spin decomposition and semiclassical expansion}
\label{a1}
With the spin decomposition and semi-classical expansion for the Wigner function $W$ and self-energies $\Sigma^<$ and $\Sigma^>$, the spin components $C_F, C_P, C_V, C_A, C_S$ and $D_F, D_P, D_V, D_A, D_S$ of the collision terms $C$ and $D$ at classical level and first order in $\hbar$ can be easily extracted from the following expressions for $\left[\Sigma^>,S^<\right]_\star$ and $\left\{\Sigma^>,S^<\right\}_\star$, 
\begin{eqnarray}
\left[\Sigma^>,S^<\right]^{(0)}_\star &=& 2i\left(\overline{\Sigma}_V^\mu A_\mu -\overline{\Sigma}_A^\mu V_\mu\right)^{(0)}i\gamma_5\nonumber\\
&& 
+2i\left(\overline{\Sigma}_P A^\mu +\overline{\Sigma}_{V\nu} S^{\nu\mu} -\overline{\Sigma}_A^\mu P
+\overline{\Sigma}_S^{\mu\nu} V_\nu\right)^{(0)}\gamma_\mu\nonumber\\
&&
+2i\left(\overline{\Sigma}_PV^\mu-\overline{\Sigma}_V^\mu P+\overline{\Sigma}_{A\nu}S^{\nu\mu}+\overline{\Sigma}_S^{\mu\nu}A_\nu\right)^{(0)}\gamma_5\gamma_\mu\nonumber\\
&&
+2i\left(\overline{\Sigma}_A^{[\mu}A^{\nu]}-\overline{\Sigma}_V^{[\mu}V^{\nu]}-\overline{\Sigma}_S^{\alpha[\mu}S^{\ \nu]}_\alpha\right)^{(0)}\sigma_{\mu\nu}/2\nonumber\\
\end{eqnarray}
and
\begin{eqnarray}
\left\{\Sigma^>,S^<\right\}^{(0)}_\star &=& 2\left(\overline{\Sigma}_FF-\overline{\Sigma}_PP+\overline{\Sigma}_V^\mu V_\mu-\overline{\Sigma}_A^\mu A_\mu+\overline{\Sigma}_S^{\mu\nu}S_{\mu\nu}/2\right)^{(0)}\nonumber\\
&&
+2\left(\overline{\Sigma}_FP+\overline{\Sigma}_PF+\epsilon_{\mu\nu\alpha\beta}\overline{\Sigma}_S^{\mu\nu}S^{\alpha\beta}/4\right)^{(0)}i\gamma_5\nonumber\\
&&
+2\left(\overline{\Sigma}_FV^\mu+\overline{\Sigma}_V^\mu F+\epsilon^{\sigma\nu\lambda\mu}\left(\overline{\Sigma}_{A\sigma} S_{\nu\lambda}+\overline{\Sigma}_{S\sigma\nu}A_\lambda\right)/2\right)^{(0)}\gamma_\mu\nonumber\\
&&
+2\left(\overline{\Sigma}_FA^\mu+\overline{\Sigma}_A^\mu F+\epsilon^{\sigma\nu\lambda\mu}\left(\overline{\Sigma}_{V\sigma} S_{\nu\lambda}+\overline{\Sigma}_{S\sigma\nu}V_\lambda\right)/2\right)^{(0)}\gamma_5\gamma_\mu\nonumber\\
&&
+2\left(\overline{\Sigma}_FS^{\mu\nu}+\overline{\Sigma}_S^{\mu\nu}F+\epsilon^{\mu\nu\alpha\beta}\left(\overline{\Sigma}_{A\alpha} V_\beta-\overline{\Sigma}_{V\alpha} A_\beta-\overline{\Sigma}_{S\alpha\beta}P/2
-\overline{\Sigma}_PS_{\alpha\beta}/2\right)\right)^{(0)}\sigma_{\mu\nu}/2
\end{eqnarray}
at classical level and 
\begin{eqnarray}
\left[\Sigma^>,S^<\right]^{(1)}_\star &=& 2i\hbar\left(\overline{\Sigma}_V^\mu A_\mu -\overline{\Sigma}_A^\mu V_\mu\right)^{(1)}i\gamma_5\nonumber\\
&&
+2i\hbar\left(\overline{\Sigma}_P A^\mu +\overline{\Sigma}_{V\nu} S^{\nu\mu}
-\overline{\Sigma}_A^\mu P+\overline{\Sigma}_S^{\mu\nu}V_\nu\right)^{(1)}\gamma_\mu\nonumber\\
&&
+2i\hbar\left(\overline{\Sigma}_PV^\mu-\overline{\Sigma}_V^\mu P+\overline{\Sigma}_{A\nu}S^{\nu\mu}+\overline{\Sigma}_S^{\mu\nu}A_\nu\right)^{(1)}\gamma_5\gamma_\mu\nonumber\\
&&
+2i\hbar\left(\overline{\Sigma}_A^{[\mu}A^{\nu]}-\overline{\Sigma}_V^{[\mu}V^{\nu]}-\overline{\Sigma}_S^{\alpha[\mu}S^{\nu]}_\alpha\right)^{(1)}\sigma_{\mu\nu}/2\nonumber\\
&&
+i\hbar\left[\overline{\Sigma}_FF-\overline{\Sigma}_PP+\overline{\Sigma}_V^\mu V_\mu-\overline{\Sigma}_A^\mu A_\mu+\overline{\Sigma}_S^{\mu\nu}S_{\mu\nu}/2\right]^{(0)}_{\text{PB}}\nonumber\\
&&
+i\hbar\left[\overline{\Sigma}_FP+\overline{\Sigma}_PF+\epsilon_{\mu\nu\alpha\beta}\overline{\Sigma}_S^{\mu\nu}S^{\alpha\beta}/4\right]^{(0)}_{\text{PB}}i\gamma_5\nonumber\\
&&
+i\hbar\left[\overline{\Sigma}_FV^\mu+\overline{\Sigma}_V^\mu F+\epsilon^{\sigma\nu\lambda\mu}(\overline{\Sigma}_{A\sigma}S_{\nu\lambda}+\overline{\Sigma}_{S\sigma\nu}A_\lambda)/2\right]^{(0)}_{\text{PB}}\gamma_\mu\nonumber\\
&&
+i\hbar\left[\overline{\Sigma}_FA^\mu+\overline{\Sigma}_A^\mu F+\epsilon^{\sigma\nu\lambda\mu}(\overline{\Sigma}_{V\sigma}S_{\nu\lambda}
+\overline{\Sigma}_{S\sigma\nu}V_\lambda)/2\right]^{(0)}_{\text{PB}}\gamma_5\gamma_\mu\nonumber\\
&&
+i\hbar\left[\overline{\Sigma}_FS^{\mu\nu}+\overline{\Sigma}_S^{\mu\nu}F+\epsilon^{\mu\nu\alpha\beta}\left(\overline{\Sigma}_{A\alpha}V_\beta
-\overline{\Sigma}_{V\alpha}A_\beta-\overline{\Sigma}_{S\alpha\beta}P/2-\overline{\Sigma}_PS_{\alpha\beta}/2\right)\right]^{(0)}_{\text{PB}}\sigma_{\mu\nu}/2
\end{eqnarray}
and  
\begin{eqnarray}
\left\{\Sigma^>,S^<\right\}^{(1)}_\star &=&2\hbar\left(\overline{\Sigma}_FF-\overline{\Sigma}_PP+\overline{\Sigma}_V^\mu V_\mu-\overline{\Sigma}_A^\mu A_\mu+\overline{\Sigma}_S^{\mu\nu}S_{\mu\nu}/2\right)^{(1)}\nonumber\\
&&
+2\hbar\left(\overline{\Sigma}_FP+\overline{\Sigma}_PF+\epsilon_{\mu\nu\alpha\beta}\overline{\Sigma}_S^{\mu\nu}S^{\alpha\beta}/4\right)^{(1)}i\gamma_5\nonumber\\
&&
+2\hbar\left(\overline{\Sigma}_SV^\mu+\overline{\Sigma}_V^\mu S+\epsilon^{\sigma\nu\lambda\mu}(\overline{\Sigma}_{A\sigma}S_{\nu\lambda}+\overline{\Sigma}_{S\sigma\nu}A_\lambda)/2\right)^{(1)}\gamma_\mu\nonumber\\
&&
+2\hbar\left(\overline{\Sigma}_FA^\mu+\overline{\Sigma}_A^\mu F+\epsilon^{\sigma\nu\lambda\mu}(\overline{\Sigma}_{V\mu}S_{\nu\lambda}
+\overline{\Sigma}_{S\sigma\nu}V_\lambda)/2\right)^{(1)}\gamma_5\gamma_\mu\nonumber\\
&&
+2\hbar\left(\overline{\Sigma}_FS^{\mu\nu}+\overline{\Sigma}_S^{\mu\nu}F+\epsilon^{\mu\nu\alpha\beta}\left(\overline{\Sigma}_{A\alpha}V_\beta
-\overline{\Sigma}_{V\alpha}A_\beta-\overline{\Sigma}_{S\alpha\beta}P/2-\overline{\Sigma}_PS_{\alpha\beta}/2\right)\right)^{(1)}\sigma_{\mu\nu}/2\nonumber\\
&&
-\hbar\left[\overline{\Sigma}_V^\mu A_\mu-\overline{\Sigma}_A^\mu V_\mu\right]^{(0)}_{\text{PB}}i\gamma_5\nonumber\\
&&
-\hbar\left[\overline{\Sigma}_PA^\mu+\overline{\Sigma}_{V\nu}S^{\nu\mu}-\overline{\Sigma}_A^\mu P+\overline{\Sigma}_S^{\mu\nu}V_\nu\right]^{(0)}_{\text{PB}}\gamma_\mu\nonumber\\
&&
-\hbar\left[\overline{\Sigma}_PV^\mu-\overline{\Sigma}_V^\mu P+\overline{\Sigma}_{A\nu}S^{\nu\mu}+\overline{\Sigma}_S^{\mu\nu}A^\nu\right]^{(0)}_{\text{PB}}\gamma_5\gamma_\mu\nonumber\\
&&
-\hbar\left[\overline{\Sigma}_A^{[\mu}A^{\nu]}-\overline{\Sigma}_V^{[\mu}V^{\nu]}-\overline{\Sigma}_S^{\alpha[\mu}S^{\nu]}_\alpha\right]^{(0)}_{\text{PB}}\sigma_{\mu\nu}/2
\end{eqnarray}
at order $\mathcal O(\hbar)$, where $(AB)^{(0)}$ and $(AB)^{(1)}$ are defined as $(AB)^{(0)}=A^{(0)}B^{(0)}$ and $(AB)^{(1)}=A^{(1)}B^{(0)}+A^{(0)}B^{(1)}$. The expressions for the commutators $\left[\Sigma^<, S^>\right]$ and $\left\{\Sigma^<, S^>\right\}$ can similarly be obtained. 
\end{appendix}

\bibliographystyle{iopart-num}
\bibliography{ref}

\end{document}